\documentclass[aps,prl,superscriptaddress,twocolumn,normalem]{revtex4-1}
\usepackage{amsmath,amssymb,bm,graphicx,epsfig,psfrag,ulem,color,natbib, comment}
\usepackage{hyperref}

\newcommand{\bs}{\mathbf {s}}
\newcommand{\br}{\mathbf {r}}
\newcommand{\bk}{\mathbf {k}}
\newcommand{\bv}{\mathbf {v}}

\begin{document}

\title{The Kelvin-wave cascade in the vortex filament model}

\author{Andrew~W.~Baggaley}
\email{andrew.baggaley@gla.ac.uk}
\affiliation{School of Mathematics and Statistics, University of
Glasgow, Glasgow, G12 8QW, United Kingdom}
\author{Jason Laurie}
\email{jason.laurie@weizmann.ac.il}
\affiliation{Department of Physics of Complex Systems, Weizmann Institute of Science, Rehovot, 76100, Israel}


\begin{abstract}
The small-scale energy transfer mechanism in zero temperature superfluid turbulence of helium$-4$ is still a widely debated topic.  Currently, the main hypothesis is that weakly nonlinear interacting Kelvin-waves (KWs) transfer energy to sufficiently small scales such that energy is dissipated as heat via phonon excitations.  Theoretically, there are at least two proposed theories for Kelvin-wave interactions. We perform the most comprehensive numerical simulation of weakly nonlinear interacting KWs to date and show, using a specially designed numerical algorithm incorporating the full Biot-Savart equation, that our results are consistent with the nonlocal six-wave KW interactions as proposed by L'vov and Nazarenko.
\end{abstract}


\maketitle

The understanding of the mechanisms involved in the transfer of energy to small scales is one of the most fundamental problems in three-dimensional classical fluid turbulence. Still, 70 years on, the theory of classical turbulence proposed by Kolmogorov and Obukhov \cite{Kolmogorov1941a,Kolmogorov1941b, Obukhov1941} forms the basis and the majority of our knowledge.  Interest in the quantum analogue, superfluid turbulence, has heightened in the last few decades, especially with experimental observation of Bose-Einstein Condensation in 1995~\cite{Anderson1995}, due to the fact that all vortices are identical and the notation that this may, in some way, provide insight into the classical theory.

In superfluids, severe quantum restrictions prevent a continuous vorticity distribution--if the superfluid is excited, any induced vorticity is defined though topological, zero density, defects within the flow.  These topological defects are known as quantized vortices. Each quantized vortex is identical--they have fixed circulation that appears as integer multiplies of $\kappa = h/m$, where $h$ is Planck's constant and $m$ is the mass of the Boson.  Moreover, the size of the vortex core, or healing length $\xi$ is extremely small, orders of magnitudes smaller than the typical inter-vortex spacing. 

Feynman defined superfluid turbulence as the study of the chaotic behavior of the superfluid flow induced by a tangle of quantized vortex lines~\cite{Feynman1955,Vinen2002,Skrbek2012}.  What makes superfluid turbulence so interesting is its similarities to classical 3D classical (Navier-Stokes) turbulence. At large scales, analogies to classical eddies can be made with polarized bundles of quantized vortex lines which invoke a large-scale flow around them~\cite{Lvov2007}. Interestingly, it has been shown experimentally~\cite{Maurer1998}, and numerically~\cite{Araki2002,Baggaley2012}, that this large-scale flows exhibits the famous Kolmogorov-Obukhov energy spectrum of classical turbulence theory. 

At small scales the similarities cease.  In classical fluid turbulence, energy is dissipated through viscosity. However, in zero temperature superfluid turbulence there is no viscosity, so how is energy dissipated? At scales of the order of the inter-vortex spacing, the polarized bundle picture breaks down, and a `cross-over' scenario is predicted, whether in the form of thermalization~\cite{Lvov2007} or through a variety of vortex interactions~\cite{Kozik2008}.  However, in both conjectures, the majority of the energy is assumed to be transferred to propagating Kelvin-waves (KWs) along quantized vortex lines.  The popular hypothesis involves energy dissipation via the excitation of phonons by high frequency KWs~\cite{Vinen2002b,Vinen2005}. These high frequency KWs are thought to be created through weakly nonlinear interactions of larger scale KWs forced through vortex reconnections at the scale of the inter-vortex spacing.  If this idea is correct, then the understanding of how KWs interact with one other becomes an important problem in the study of energy transfer in superfluid turbulence.  We note, that other possible energy transfer mechanisms may exist, such as Feynman's picture of a cascade of vortex rings~\cite{Feynman1955}, or via the emission of vortex rings at reconnection~\cite{Nemirovskii2010,Kondaurova2012}. However, Svistunov argued Feynman's scenario breaks energy conservation of the quantum tangle~\cite{Svistunov1995}, more recently this assertion was challenged by Nemirovskii \cite{Nemirovskii2013}. Nevertheless the formation of vortex rings at reconnection events only occurs for small reconnection angles~\cite{Kursa2011}. We take the view that dissipation due to KWs will be particularly important in semi-classical (quasi-Kolmogorov) superfluid turbulence where polarization of the vortex lines means reconnection angles are typically small. However, dissipation due to reconnections and loop emission is probably crucial in the random, unpolarized, tangle \cite{Lvov2007}.

Nonetheless,  the main topic of this manuscript is in KW interactions at scales far smaller than the inter-vortex spacing, where the potential cross-over mechanisms are assumed to be irrelevant. The natural approach to this problem is to utilize the wave turbulence theory paradigm~\cite{Zakharov1992, Nazarenko2011}, which has lead to several attempts to statistically describe weakly nonlinear KW interactions based on the notion of a constant energy flux transfer to high frequency KWs~\cite{Kozik2004, Boffetta2009, Lvov2010}.  Unfortunately, none of the theories have been universally accepted within the community because of mathematical technicalities and poor numerical evidence.
  
The main goal of this manuscript is therefore to provide indisputable numerical evidence for a particular KW theory.  To achieve this, we develop a new numerical scheme,  based upon the vortex filament method of Schwarz~\cite{Schwarz1988}.  We show that using this approach, we are able to distinguish between the competing KW theories and show significant indication for the validity of a particular one.

The understanding of KW interactions is not a new topic, there has been many theoretical~\cite{Lebedev2010a,Lebedev2010b, Kozik2010a, Kozik2010b,Sonin2012,Sonin2012b,Lvov2012} and numerical~\cite{Vinen2003,Kozik2005,Boffetta2009,Boue2011, Kozik2010arxiv, Hanninen2011,Krstulovic2012} works for the two main KW theories.  In the most basic of terms, one is based on the assumption of local six-wave interactions~\cite{Kozik2004}, whilst the other on non-local six-wave interactions leading to local four-wave processes~\cite{Laurie2010,Lvov2010}.  The reason we feel the need to add to these numerical contributions, is because none have been of sufficient calibre to satisfy both camps. We feel that we have carefully considered all the criticisms of the previous vortex filament model simulations of weakly nonlinear interacting KWs and believe to have developed an algorithm will be able to address all of the previous critiques.


Before we outline the numerical procedure, we first summarize the key theoretical results of the two KW theories.  Both begin with the idealized consideration of a single periodic (in $z$) quantized vortex line, approximated by a one-dimensional (1D) space curve ${\bs}(\zeta,t)$ in a three-dimensional domain, evolved via the Biot-Savart equation:
\begin{equation}
\dot{\bs}=  {\bv_{\rm si}}=\frac{\kappa}{4 \pi} \oint_{\cal L} \frac{(\bs-\br) }
{\vert \bs - \br \vert^3}
\times {\bf \rm d}\br.
\label{eq:BS}
\end{equation}  
The integral is taken over the entire vortex configuration $\cal L$  with $\kappa=9.97 \times 10^{-4}~\rm cm^2/s$. 
The Biot-Savart equation~\eqref{eq:BS} contains a singularity as ${\br \to \bs}$, which is usually circumvented by the introduction of a strict cut-off $|\bs - \br|> \xi$, where  $\xi$ is taken to be the vortex core radius of superfluid helium$-4$: $\xi= 1\times 10^{-8}~{\rm cm}$.



\begin{figure}
\begin{center}
\includegraphics[width=0.5\columnwidth]{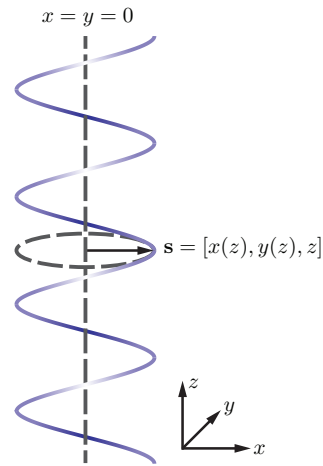} 
\caption{Schematic of the periodic vortex line in $z$.}
\label{fig:vortex}
\end{center}
\end{figure}

It was shown in~\cite{Svistunov1995} that by considering a single periodic quantized vortex line directed along the $z$-axis, with small perturbations with respect to the straight line configuration, such that the vortex position can be represented as ${\bs}=(x(z),y(z),z)$ (see Fig.~\ref{fig:vortex}), and if the single-valuedness of the functions $x(\cdot)$ and $y(\cdot)$ are preserved such that the vortex line cannot fold upon itself, then the Biot-Savart equation can be represented in Hamiltonian form $i \kappa\dot{a} = \delta \mathcal{H}[a]/\delta a^*$ for the complex canonical coordinate $a(z,t)=x(z,t)+i y(z,t)$ with Hamiltonian:
\begin{equation}
\mathcal{H}[a] = {\kappa^2 \over 4 \pi} \int  {1 + {\rm Re}(a'^{*}(z_1) a'(z_2))
\over \sqrt{(z_1 - z_2)^2 + |a(z_1)-a(z_2)|^2}}\, {\rm d}z_1\, {\rm d}z_2,
\label{eq:h2d}
\end{equation}
where $a'(z) = \partial a/\partial z$.  For small perturbations $|a'(z)| \ll 1$, which is precisely the regime for weakly nonlinear KWs, one can expand Hamiltonian \eqref{eq:h2d} in terms of wave amplitudes $a_\bk(t)$, where $a(z,t) = \kappa^{-1/2}\sum_\bk a_\bk(t)\exp(i\bk z)$, here $\bk\in\mathbb{R}$ and $k=|\bk|$.  The leading contribution to the Hamiltonian describes the linear evolution of KWs with dispersion relation
\begin{equation}\label{eq:dispersion}
\omega(\bk)=\omega_k = \frac{\kappa k^2}{4\pi}\left[\ln\left( \frac{1}{k\xi}\right) -\gamma -\frac{3}{2}\right],
\end{equation}
where $\gamma=0.5772$ is the Euler-Mascheroni constant.  The exact value of the constant in Eq.~\eqref{eq:dispersion} is dependent on the vortex core shape.  Our model uses the Biot-Savart cut-off, but others have been considered, such as a hollow core~\cite{Barenghi2006} or a uniform vorticity distribution~\cite{Schwarz1985}, each giving their own value.

 The next order contributions correspond to four-wave interactions followed by a six-wave term. Detailed derivations of the interaction coefficients of the Hamiltonian can be found in~\cite{Laurie2010}.   Once in this form, one can apply the wave turbulence formulation~\cite{Zakharov1992, Nazarenko2011}.  This system does not contain any resonant four-wave interactions, i.e. nontrivial solutions of the four-wave resonant condition: $\bk+\bk_2=\bk_3 + \bk_4$ and $\omega_k + \omega_2 = \omega_3 +\omega_4$ (where $\omega_i=\omega(\bk_i)$) because of the form of the dispersion relation~\eqref{eq:dispersion}.  Therefore, one has to perform a quasi-linear canonical transformation to the wave action variable $a_\bk$ to remove the non-resonant four-wave interactions which subsequently lead to the main nonlinear contribution being of order six~\cite{Laurie2010}.   Assuming local KWs interactions, i.e. that the only resonant wave interactions arise from waves with a similar magnitude of wavenumber $\bk$, wave turbulence theory predicts a six-wave kinetic equation $\dot{n}_\bk = {\rm St}_6[n_\bk]$ (where the detailed expression for the six-wave collision integral ${\rm St}_6[n_{\bk}]$ is given by Eq.~(8) of~\cite{Kozik2004}) for the evolution of the wave action density $n(\bk,t)=n_\bk=\langle |a_\bk|^2\rangle$, where the average $\langle \cdot\rangle$ is taken over an ensemble of realizations.  An amendment to the collision integral ${\rm St}_6[n_\bk]$, taking into account sub-leading corrections to the resonant manifold condition can be found in~\cite{Laurie2010}.  This kinetic equation contains two non-equilibrium steady state power-law solutions ($\dot{n}_\bk=0$), known as Kolmogorov-Zakharov spectra, corresponding to a constant flux cascades of conserved quantities. For us, we are interested in the one corresponding to the constant flux of energy, $\epsilon_k$, to small scales, which expressed in terms of the energy spectrum $E_k = \omega_k\left(n_{\bk}+n_{-\bk}\right)$ for $\bk >0$ is
 \begin{subequations}
 \begin{equation}\label{eq:KS}
E_k = C_{\rm KS}\,\kappa^{7/5}\Lambda_k\,\epsilon_k^{1/5}k^{-17/5},
\end{equation}
where $\Lambda_k=\ln\left(1/k\xi\right)-\gamma-3/2$.  The spectrum~\eqref{eq:KS} is known as the Kozik-Svistunov (KS) spectrum with dimensionless prefactor $C_{\rm KS}$.   However, it was shown in~\cite{Laurie2010} that the locality of interaction assumption used to derive~\eqref{eq:KS} was incorrect, making the KS spectrum invalid.  Subsequently, a nonlocal theory was proposed by L'vov and Nazarenko and an alternative local four-wave kinetic equation ($\dot{n}_\bk={\rm St}_4[n_\bk]$, see Eq.~(5a) of~\cite{Lvov2010}) was derived for weakly interacting KWs upon curved vortex lines.  In a similar way, the four-wave kinetic equation exhibits a non-equilibrium stationary solution that corresponds to a constant energy flux.  This solution is known as the L'vov-Nazarenko spectrum and is given by
 \begin{equation}\label{eq:LN}
E_k = C_{\rm LN}\,\kappa\,\Lambda_k\,\epsilon_k^{1/3}\,\Psi^{-2/3} k^{-5/3},
\end{equation}
with dimensionless prefactor $C_{\rm LN}$ and $\Psi=(2/\kappa)\sum_{k} k^2\,n_k$.  The most na\"{i}ve way to differentiate between the two theories is to compare the power-law behaviors of the wave action spectra~\eqref{eq:KS} and~\eqref{eq:LN}. However, their power-law exponents are very close, making any distinguishable observation extremely difficult.  Alternatively, one can try to measure the dimensionless prefactors $C_{\rm KS}$ and $C_{\rm LN}$, which should be of order one if the spectrum is realizable.  An attempted was made in this direction in~\cite{Boue2011} for the Local Nonlinear equation, (a partial differential equation derived directly from the Biot-Savart equation in the nonlocal KW limit) where $C_{\rm LN}= 0.304$ was successfully obtained through mathematical arguments and numerically verified. On the other hand, efforts of a similar nature applied to the full Biot-Savart equation have, as of yet, been unsuccessful. 

When the inclination of the KW slope approaches $|a'(z)| \sim 1$, then the weakly nonlinear hypothesis of wave turbulence theory fails and the linear timescale of the wave motion $\tau_L = 2\pi/\omega_k$ becomes the order of the nonlinear timescale determined through the kinetic equation $\tau_{NL} = \partial \ln(n_k)/\partial t$ over a wide range of scales.  This criterion is precisely the critical balance condition for strong wave turbulence.  By requiring that the linear and nonlinear timescale match across all spatial scales, one can derive a critical balance energy spectrum, originally derived by Vinen~\cite{Vinen2005}, which is given as
\begin{equation}\label{eq:CB}
E_k = C_{\rm CB}\,\kappa^2 k^{-1},
\end{equation}
\end{subequations}
Interestingly, the power-law exponent of~\eqref{eq:CB} coincides with the second non-equilibrium Kolmogorov-Zakharov spectrum for the constant flux of wave action to large-scales which occurs because the total wave action $\mathcal{N}[a]=\int_{\mathcal{D}}|a(z)|^2 \, {\rm d}z$ is another conserved quantity of the Biot-Savart dynamics~\eqref{eq:BS}.  We note that although the wavenumber exponent maybe the same, the dimensional prefactors and constant will be different.

Spectra~\eqref{eq:KS},~\eqref{eq:LN} and~\eqref{eq:CB} are three possible power-law scalings for the cascade of energy to small-scales, each with their own physical justifications. To discover which, if any, are realizable, and therefore which physical mechanisms are important for KW interactions we perform a numerical simulation of weakly nonlinear interacting KWs.

We discretize the Biot-Savart equation~\eqref{eq:BS} following the numerical method proposed by Schwarz \cite{Schwarz1988}, namely the vortex filament model, where quantized vortices are modeled as 1D vortex filaments with fixed circulation $\kappa$.  The self-induced velocity $\bv_{\rm si}$ of the vortex filaments is given by the Biot-Savart equation~\eqref{eq:BS}, de-singularized in a standard way, using the local induction approximation to take into account contributions from neighboring points~\cite{Schwarz1988}.

  In order to sustain a non-equilibrium steady state of KWs on the vortex line we include forcing and dissipation. There are several strategies one can take in forcing KWs onto the vortex line, such as a continual vibration of the vortex line \cite{Vinen2003,Hanninen2011} or by additively exciting a specific range of KWs from rest \cite{Boffetta2009,Boue2011}.   In our case, we implement the latter scenario.  The forcing is incorporated by the addition of term to the right-hand side of~\eqref{eq:BS}, $\bv_{\rm f}$, that adds a collection of randomly orientated KWs at a specific scale at each time step. The forcing is situated at large-scales in a narrow annulus around the tenth harmonic.  This is to ensure that we induce a direct energy cascade with an inertial range as large as possible. In the forcing region we add KWs with fixed amplitude ${\cal A}$, and with random phases $\phi_\bk$ uniformly distributed in  $[0,2\pi)$ for each Fourier mode and at each time step.  The forcing term takes the form of
\begin{eqnarray}
\bv_{\rm f}&=& \left[{\rm Re}(f),{\rm Im}(f),0\right],\nonumber\\ {\rm where} \quad f&=&\sum_{9\leq k \leq 11} {\cal A} \exp\left(i\bk  \, z+i\phi_\bk\right).
\label{eq:forcing}
\end{eqnarray}
In reality, KWs are forced through sporadic and random vortex reconnections that do not simply excite KWs at a specific scale.  In this regard we have made a simplification.  However, at scales far smaller than the inter-vortex spacing, we expect that the KWs do not sense the reconnection event and that the simplification is justified.
 
To model the eventual energy dissipation via phonons and to prevent a large-scale bottleneck, we apply an exponential filter, $D[\cdot]$, at each time step that acts primarily in the low and high wavenumber regions of Fourier space.  To apply the filter we are resorted to re-meshing the vortex line after every time step onto a uniform grid in $z$ with constant grid spacing $\Delta z = 2\pi/1024$ using a cubic spline.  This permits us to apply the Fast Fourier Transform to the perturbations around the initial straight line configuration $x=y=0$.  This gives the Fourier amplitudes $a_\bk$, to which we can apply the filter given by 
\begin{equation}\label{eq:D}
D[a_\bk] = \begin{cases} e^{-\Delta t\left(\alpha k^{-2} + \nu k^4 \right)}a_\bk &\mbox{if } k\neq 0. \\ 
a_\bk = 0 & k=0, \end{cases}
\end{equation}
This removes energy located in the infra-red and ultra-violet regions of Fourier space~\footnote{We remark that low wavenumber dissipation is required in order to prevent the formation of a strongly nonlinear condensate at the largest-scale through the inverse cascade of wave action.}.  We inverse Fast Fourier Transform back to physical space to give the post-filtered vortex line on the uniform grid.  Subsequently, the forcing term is applied and the vortex line in evolved by the Biot-Savart equation~\eqref{eq:BS} using a third order Runge-Kutta time stepping scheme. 

\begin{figure}
\begin{center}
\includegraphics[width=\columnwidth]{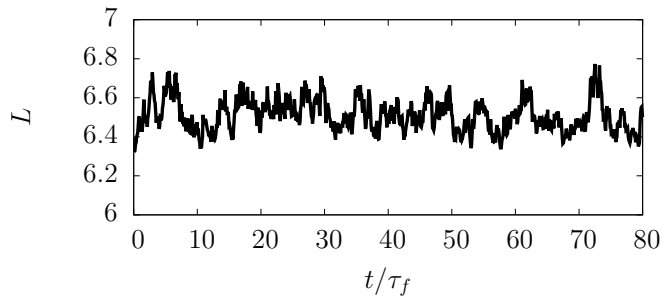} 
\caption{
The time evolution of the total vortex line length $L=\int_{\cal L} d\zeta$, which saturates indicating that a non-equilibrium statistical steady state has been reached. Time has been nondimensionalised using the linear timescale at the forcing scale defined as $\tau_f=2\pi/\omega_{k=10}\simeq 56.402$.}
\label{fig:linelength}
\end{center}
\end{figure}

\begin{figure}
\begin{center}
\includegraphics[width=\columnwidth]{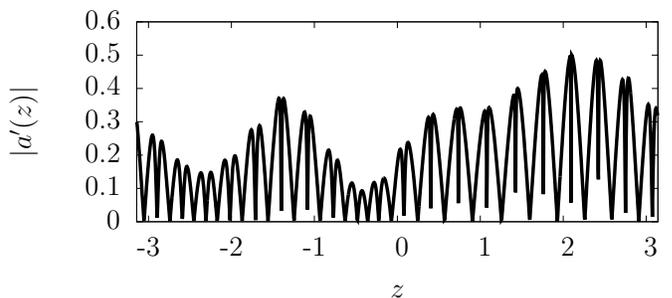} 
\caption{
Magnitude of the derivative of the wave amplitude $|a'(z)|$ for the weakly driven simulation.  We observe that the KW condition $|a'(z)|\ll 1$ is satisfied throughout $z$.}
\label{fig:wprime}
\end{center}
\end{figure}

 
We perform the numerical simulation in a box that is open in $x$ and $y$, whilst periodic in $z$ of length ${\cal D}=2\pi~ \rm cm$.  The initial condition is a single unperturbed (straight) vortex line along the $z$ direction in the center of the box ($x=y=0$).  We discretize the vortex into $1024$ uniformly spaced grid points. We invoke periodic boundary conditions in $z$, and the simulations are progressed in time with a fixed time step $\Delta t=1 \times 10^{-3}\,\rm s$, performing $8 \times 10^{6}$ time steps. For the exponential filter, we use the following parameters: $\alpha =  1\times 10^{-1}~\rm cm^2/s$ and $\nu = 1\times 10^{-9}~\rm cm^4/s$.
The forcing amplitude is chosen to be $\mathcal{A}=0.05$ cm/s, to ensure that the system remains weakly nonlinear, whilst strong enough to generate nonlinear KW interactions.   The simulation is evolved so that the system reaches a non-equilibrium stationary state indicated by statistical stationarity of the total vortex line length of the tangle as shown in Fig.~\ref{fig:linelength}.  Moreover, in Fig.~\ref{fig:wprime}, we present a typical snapshot of the magnitude of $a'(z)$ along the vortex line. We observe that the value is $|a'(z)|\ll1$ along all the line, verifying the weak KW condition for the weak wave turbulence theory.  



\begin{figure}
\begin{center}
\includegraphics[width=\columnwidth]{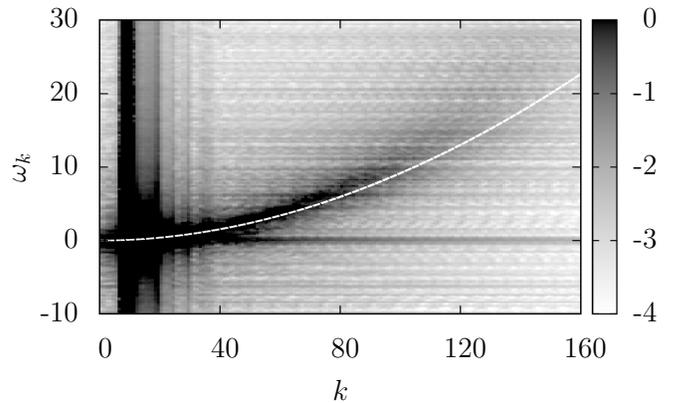}         
\caption{
$(k,\omega_k)$-plot of the logarithm of the intensity of the 2D Fourier transform of the wave amplitude $a(z,t)$.  The forcing region is indicated by the vertical black structure around $k=10$, whilst the dispersion curve for weakly nonlinear KWs can be observed emanating from the forcing region.. The white dashed curve depicts the theoretical dispersion relation~\eqref{eq:dispersion} of linear propagating KWs.}
\label{fig:dispersion}
\end{center}
\end{figure}

An additional check can be performed by taking a 2D Fourier transform of the wave amplitude $a(z,t)$ in space and time.  By plotting the intensity of the Fourier amplitudes, one can probe the dispersion curve of propagating wave. In Fig.~\ref{fig:dispersion}, we present this $(k,\omega_k)-$plot and observe a dispersion curve that is almost in perfect agreement with the theoretical result for linear propagating KWs, Eq.~\eqref{eq:dispersion}.  The slight vertical shift is most likely a consequence of the weak nonlinearity, and the subsequent frequency shift associated to this. The vertical structure we observe across all values of $\omega_k$ around $k=10$ is certainly from the additive forcing scheme of our numerical setup.

\begin{figure}
\begin{center}

\includegraphics[width=\columnwidth]{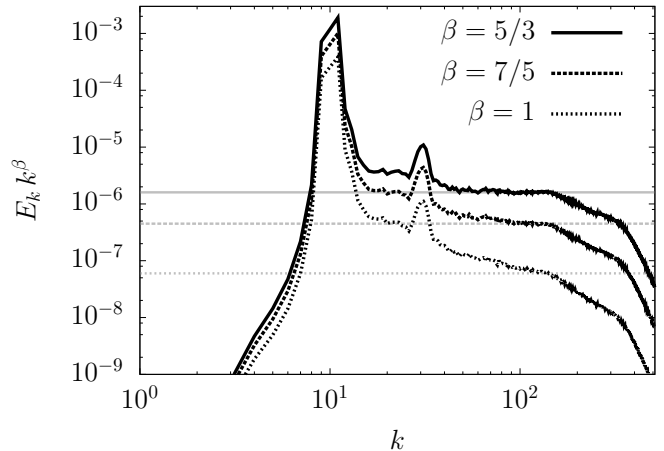}         
\caption{
 The compensated wave action spectra with $\mathcal{A} =0.05$ cm/s: compensated by $E_k\, k^{5/3}$, Eq.~\eqref{eq:LN} (solid line), $E_k \,k^{7/5}$, Eq.~\eqref{eq:KS} (dashed line), and finally $E_k\, k$, Eq.~\eqref{eq:CB}  (dotted line). The grey horizontal lines indicate flat compensated spectra.}
\label{fig:spectrum}
\end{center}
\end{figure}

 To measure the energy spectrum $E_ k$ of the vortex line, we readily construct the complex canonical position $a(z)=x(z)+iy(z)$ on the uniform mesh, and hence obtain the wave action density $n_k$ by utilizing the Fast Fourier Transform and averaging over a sufficiently long time window to ensure a low noise spectrum.  Finally by multiplication by the KW frequency $\omega_k$ given by Eq.~\eqref{eq:dispersion} we obtain $E_k$ .  In Fig.~\ref{fig:spectrum} we plot the averaged energy spectrum $E_k$ compensated by the power-law spectra of \eqref{eq:KS}, \eqref{eq:LN} and \eqref{eq:CB}.  We observe excellent agreement to the L'vov-Nazarenko spectrum~\eqref{eq:LN} for almost a decade in wavenumber space.  The forcing scale can clearly be seen by the peak at small wavenumbers.  We emphasize that this result was achieved from modeling the full Biot-Savart equation \eqref{eq:BS} and represents a true non-equilibrium steady state starting from a straight vortex line, unlike the results from~\cite{Kozik2005,Boffetta2009,Boue2011}. We justify the use of a modest number of grid points in order to obtain a large quantity of data in time (almost $60$ forcing turnover times, $\tau_f$) to average out the inherent fluctuations in the energy spectrum $E_k$.
 
\begin{figure}
\begin{center}

\includegraphics[width=\columnwidth]{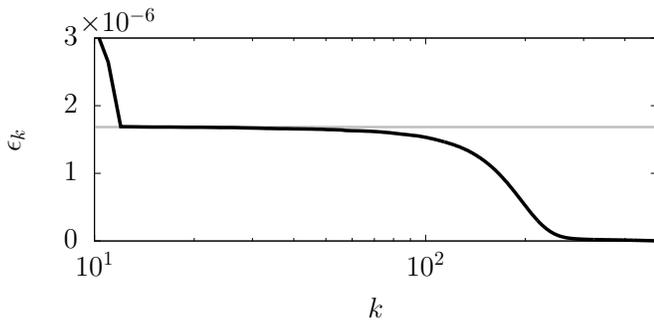}         
\caption{
Plot of the energy flux $\epsilon_k$ given by Eq.~\eqref{eq:flux} along verses $k$.  We observe a plateau of constant energy flux in the inertial range.  The straight grey line indicates the values $\epsilon_k=1.685\times 10^{-6}~\rm cm^4/s^3$.}
\label{fig:flux}
\end{center}
\end{figure}

 Finally, to make a clear distinction between the possible theories, we estimate the numerical prefactor of energy spectrum $E_k$.  To achieve this, we are required measure the energy flux, $\epsilon_k$, to small scales.  Due to the statistical stationarity of the simulation, we approximate the energy flux by measuring the energy dissipation rate at small scales from the dissipation operator $D[\cdot]$ defined in Eq.~\eqref{eq:D}.  From the wave turbulence definition of the energy flux:
 \begin{equation}
 \frac{\partial E_k}{\partial t} +\frac{\partial \epsilon_k}{\partial k}=0,
 \end{equation}
we compute the temporal derivative of $E_k$, using the fact that $D[\cdot]$ is applied at each time step. Then, the energy flux is given by
 \begin{equation}\label{eq:flux}
 \epsilon_k = \left\langle  2\sum_{k'=k_{\rm max}}^{k} \left(\alpha k^{-2} + \nu k^4\right)\omega_{k'}\left(|a_\bk|^2+|a_{-\bk}|^2\right) \right\rangle,
 \end{equation} 
where $\langle\cdot\rangle$ denotes a time average over several linear timescales at the forcing scale.  In Fig.~\ref{fig:flux} we plot the energy flux from the forcing scale $k\sim10$.  We observe a constant plateau for almost a decade in wavenumbers before the small-scale dissipation acts.  The observed peak around the forcing scale is due to the low wavenumber dissipation from $D[\cdot]$ and the fact that we neglected the contribution from the forcing.  By measuring the energy flux, we are able, from Eq.~\eqref{eq:LN}, to estimate the constant $C_{\rm LN}$.  By assuming that the energy spectrum is of the form $E_k=Ck^{-5/3}$, we observe from Fig.~\ref{fig:spectrum} that  $C=1.6\times 10^{-6}$ (the horizontal grey solid line).  Moreover, we can compute the value of $\Psi$, defined earlier, to be $\Psi = 145.622$.  Finally with using the value of the energy flux in the inertial range, $\epsilon_k=1.685\times 10^{-6}~\rm cm^4/s^3$, we estimate $C_{\rm LN}\simeq 0.318$.  This value is in remarkable agreement with the analytical prediction of $C_{\rm LN}=0.304$ from~\cite{Boue2011}, which is within $5\%$.  By contrast, if we assume that the energy spectrum $E_k$, is of Kozik-Svistunov form (Eq.~\eqref{eq:KS}), then from Fig.~\ref{fig:spectrum}, $C=4.5\times 10^{-7}$ (the horizontal grey dashed line), leading to an estimate of $C_{\rm KS}\simeq 8.7\times 10^{-3}$, which is almost two orders of magnitude smaller and clearly not order one. Indeed, this leads to the conclusion that the Kozik-Svistunov spectrum is not physically realizable. For completeness, applying a similar argument to the critical balance spectrum~\eqref{eq:CB}, where we make the crude approximation that $C=6.0\times 10^{-8}$ (the horizontal grey dotted line), we get the prefactor to be $C_{\rm CB}\simeq 6.0\times 10^{-2}$, which is again small.

To conclude, we have presented a numerical simulations of a single periodic vortex line forced from rest, modeled by the Biot-Savart equation.  We have made no local/nonlocal approximations to the equation of motion, and have considered a regime that is forced/dissipated with the formation of a non-equilibrium stationary state. We observe clear agreement to the L'vov-Nazarenko spectrum and estimated the order one prefactor to within $5\%$ of the theoretical prediction.  Additionally, we have shown that the alternative, Kozik-Svistunov theory leads to an incredibly small energy spectrum prefactor that indicates that the spectrum is not realizable.  In summary,  all this evidence indicates that weakly nonlinear KW interactions are governed by the nonlocal wave turbulence theory proposed in \cite{Lvov2010}. 


We would like to acknowledge the valuable comments of Carlo Barenghi and Sergey Nazarenko. We also acknowledge the support of the Carnegie Trust and the ANR program STATOCEAN (ANR-09-SYSC-014).

\end{document}